\documentclass[lettersize,journal]{IEEEtran}
\IEEEoverridecommandlockouts

\usepackage{cite}
\usepackage{amsmath,amssymb,amsfonts}
\usepackage{algorithmic}
\usepackage{graphicx}
\usepackage{textcomp}
\usepackage{xcolor}
\usepackage{float}
\usepackage{flushend}       
\usepackage{microtype}       
\usepackage[T1]{fontenc}     


\def\BibTeX{{\rm B\kern-.05em{\sc i\kern-.025em b}\kern-.08em
    T\kern-.1667em\lower.7ex\hbox{E}\kern-.125emX}}
\usepackage{tikz}
\usetikzlibrary{shapes.geometric,shapes.misc,arrows.meta,calc,fit,backgrounds}

\begin{document}

\title{Sub-Band Full Duplex Resource Allocation: A Predictive Deep Reinforcement Learning Approach \\

}
\author{
    Abhiram D, 
    Aiswarya Rajan, 
    Arin Shemeem, 
    Vipindev Adat Vasudevan, 
    Abdulla P
    
\thanks{Abhiram D*, Aiswarya Rajan, Arin Shemeem, and Abdulla P are with the School of Engineering, Cochin University of Science and Technology, Cochin, Kerala, India and Vipindev Adat Vasudevan is with Adat Nextwave Consulting, Kerala, India and Global Young Researchers' Academy (GYRA), Oregon, USA.
(*abhiramd010000@gmail.com).

This work has been submitted to the IEEE for possible publication. Copyright may be transferred without notice, after which this version may no longer be accessible.}%

}

\maketitle

\begin{abstract}
This paper presents a predictive deep learning framework for dynamic sub-band allocation in Sub-Band Full Duplex (SBFD) systems, addressing the challenge of balancing uplink (UL) and downlink (DL) performance under highly dynamic traffic conditions. The key contribution lies in integrating a hybrid Bidirectional Long Short-Term Memory (Bi-LSTM) model for traffic forecasting with a Double Deep Q-Network (DDQN) for real-time resource allocation. Using both predicted traffic and current queue states, the proposed system enables proactive scheduling based on traffic demand. Evaluation results show that the prediction model achieves high accuracy in capturing bursty traffic patterns, while the DDQN agent effectively adapts UL/DL split ratios according to traffic variations. The framework improves spectrum utilization, reduces queue buildup, and avoids inefficient static configurations. The proposed approach demonstrates that combining predictive intelligence with reinforcement learning significantly enhances the efficiency and adaptability of SBFD systems, making it a strong candidate for autonomous resource management in future 6G networks.
\end{abstract}

\begin{IEEEkeywords}
Sub-Band Full Duplex, Resource allocation, Traffic Forecasting, Predictive deep reinforcement learning   
\end{IEEEkeywords}

\section{Introduction}
6G networks are evolving to meet the massive data demands of immersive applications such as augmented reality (AR) \cite{10322917}, virtual reality (VR) \cite{11119950}, and smart cities. It builds on 5G by integrating artificial intelligence (AI) and machine learning across all layers of the network. Ultimately, 6G is expected to achieve ultra-high data rates of up to 1 Tbps, extremely low latency below 0.1 milliseconds, two-to-three times the spectrum efficiency, and up to ten times the energy efficiency of its predecessor \cite{jiang2021road}.

To enable these advanced applications, multiple enhancements in the Radio Access Network (RAN) are considered. In this work, we focus on SBFD as a key 6G enabler. SBFD helps lower networking latency compared to legacy 5G Time Division Duplex (TDD) \cite{wei2023performance}, which typically favours DL transmission and limits the UL. Using the SBFD mode, more UL slots can be allocated dynamically without compromising DL performance.

The greatest challenge in deploying SBFD is managing the split allocation to optimally balance the trade-off between UL and DL based on traffic demand \cite{chen2024sbfd_tradeoffs}. Existing literature and early SBFD evaluations demonstrate that configuring specific symbols for SBFD while maintaining others as standard UL or DL can effectively prioritise high-priority traffic \cite{li2024_3gpp_sbfd}. To mitigate inter-subband leakage, current frameworks rely on frequency splits implemented with guard resource blocks (RBs), inserted between the active sub-bands.

Native AI is envisioned as a foundational block of 6G architecture \cite{hexax2023vision} to enable smart, autonomous communication systems; its full potential for dynamic, predictive sub-band allocation in SBFD frameworks remains underexplored. In this paper, we address this gap by proposing a Bi-LSTM network inspired by the broader spatio-temporal traffic modeling principles used in \cite{9005997}, paired with a DDQN \cite{iqbal2021ddqn}, a reinforcement learning (RL) agent, to dynamically allocate the UL sub-band size based on predicted traffic demands. By predicting traffic bursts before they occur, our model ensures that the required UL resources are provided with minimal latency, while preventing unnecessary degradation of DL performance. Our main contributions are as follows:

\textbf{1. Predictive AI-Driven Split Allocation}: AI is used to determine the exact split that should be provided for the UL traffic inside a DL sub-band, ensuring that required UL resources are provided without degrading DL performance.

\textbf{2. Hybrid Bi-LSTM Forecasting}: To forecast future network states, we propose a hybrid 1D Convolutional Neural Network (1D-CNN) and Bi-LSTM architecture. This enables multistep time series forecasting to faithfully capture the time-varying intensity of network demand and predict rapid load fluctuations.

\textbf{3. DDQN-Based Dynamic Scheduling:} The proposed work addresses dynamic UL scheduling in an SBFD system using an RL framework combined with traffic prediction. The problem is modeled as a Markov Decision Process (MDP) \cite{bellman1957markovian}, where the DDQN agent makes scheduling decisions on a slot-by-slot basis.

\section{Sub-band Full Duplex}

\begin{figure}[!t]
    \centering
    \includegraphics[width=\columnwidth]{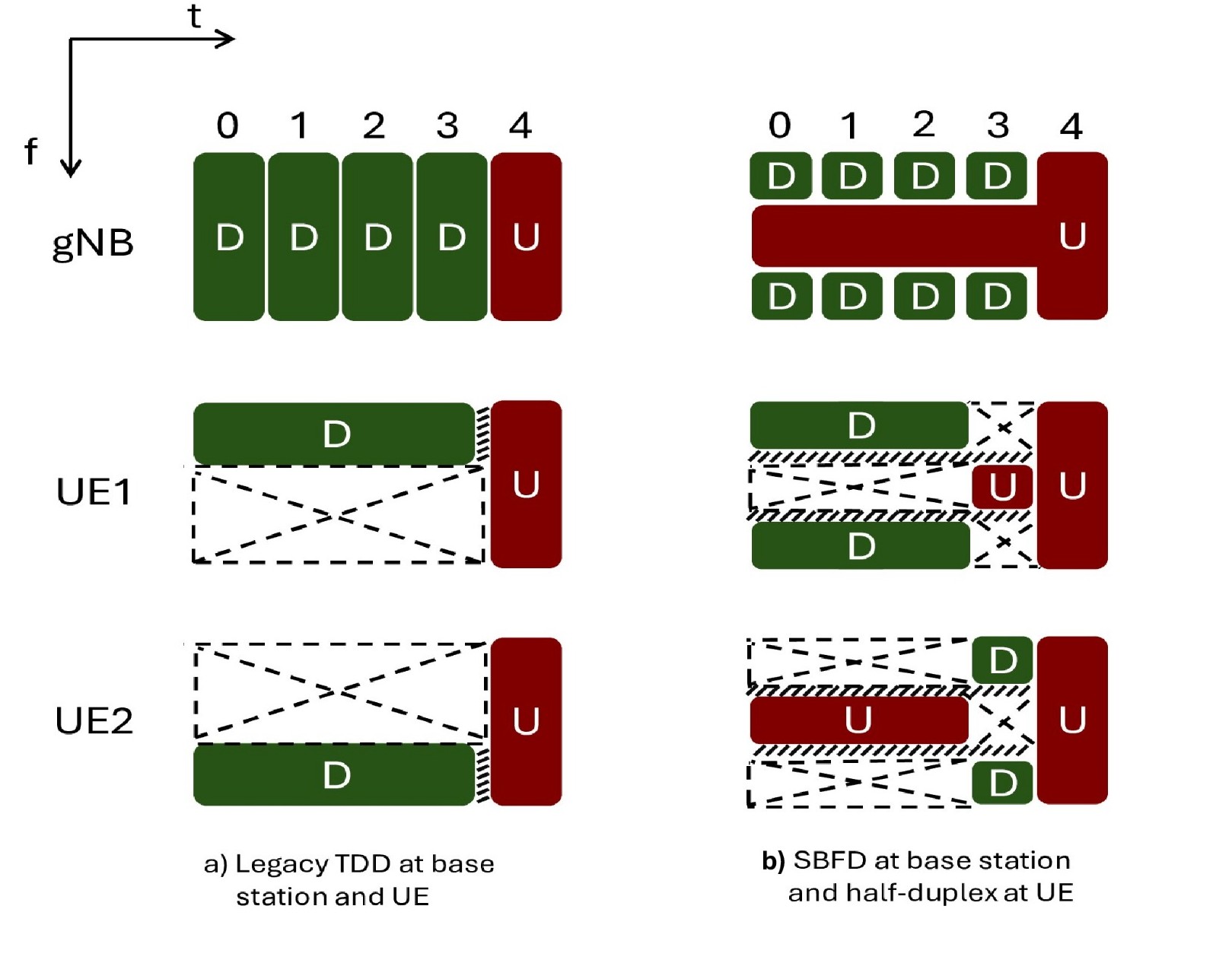}
    \caption{Slot split in TDD and SBFD}
    \label{fig:slot_split_new}
\end{figure}

The key objective of SBFD is to enhance UL coverage and reduce UL latency, particularly for power-limited User Equipment (UEs). SBFD enables the simultaneous transmission of DL and reception of UL on the same TDD carrier, within a single configured DL and UL Bandwidth Part (BWP) pair with aligned center frequencies \cite{wei2023performance}. In this architecture, full duplexing mode is deployed exclusively at the next-generation NodeB (gNB), allowing it to receive UL signals within a transmitting DL sub-band.
Meanwhile, the UEs continue to operate in half-duplex mode under the control of the Medium Access Control (MAC) layer scheduler, allowing SBFD-aware and legacy UEs to coexist seamlessly within the same cell.
The SBFD system partitions resources for DL and UL in both the time domain and frequency domain, providing flexibility for adaptation to different types of traffic. In the SBFD pattern, the DL frequency is split, separated by a guard band, and UL is sandwiched between two DL resource blocks, such as ``DUD'', within a slot.
Based on the latest 5G specification \cite{li2024_3gpp_sbfd}, almost all the bands above 3 GHz are designated as TDD bands for new radio (NR). Fig.~\ref{fig:slot_split_new} demonstrates how the split occurs in an SBFD pattern in comparison with the TDD slot configuration \cite{10404101}.

In every TDD symbol, a frequency split is implemented to facilitate SBFD. A single SBFD symbol in a TDD carrier is split in the frequency domain into one UL sub-band and one or two DL sub-bands \cite{10404101}. To prevent the leakage of a strong DL signal into the UL slot, empty guard RBs are inserted between the sub-bands. The base station signals the number of guard RBs and the duration of the UL sub-band via Radio Resource Control (RRC).

While the SBFD approach significantly increases UL capacity and spectrum efficiency, it introduces critical trade-offs and challenges \cite{chen2024sbfd_tradeoffs}. Because UL and DL share the same fixed bandwidth, widening the UL sub-band inherently narrows the DL sub-band, which can degrade performance during high DL traffic loads. Deploying full-duplex technology at the gNB causes severe interference, notably Self-Interference (SI), where powerful DL signals leak into the UL sub-band.
This architecture creates Cross-Link Interference (CLI) from the UE perspective, primarily Co-channel UE-UE CLI and Adjacent-channel UE-UE CLI \cite{han2022interference}.

The challenge in an SBFD framework is carefully trading off DL performance loss with improvements in UL performance. Legacy systems rely on static or semi-static RRC configurations, which cannot adapt to rapid, bursty traffic changes.

\section{Proposed System Model for Dynamic Sub-Band Allocation in SBFD slots}
This section details the two-stage architecture from Section I: a hybrid 1D-CNN and Bi-LSTM for traffic forecasting, followed by a DDQN agent for real-time sub-band allocation. We model the highly bursty and unpredictable behaviour of real-world 6G traffic using a Markov-Modulated Poisson Process (MMPP) \cite{el2024markov}. While this accurately simulates the traffic environment, the scheduling limitations identified in Section II motivate the need
 for a predictive approach. The Bi-LSTM serves as a forecast mechanism to overcome this limitation, providing the RL agent with advanced knowledge of impending demand shifts.
The proposed architecture begins with a time-series prediction module, wherein the 1D-CNN and Bi-LSTM network ingests a sliding window of past traffic intensities and outputs a multi-step forecast of UL and DL demands. This predicted traffic vector is then formulated
as the environmental state for the RL agent. For determining the optimal UL/DL resource split ratio, a DDQN agent is used. The DDQN decouples action selection from target Q-value evaluation, ensuring stable convergence under non-stationary traffic conditions.

\subsection{Bi-LSTM}
Our prediction architecture processes the traffic data in two specialized stages. First, the 1D-CNN layer utilizes a sliding filter with a kernel size of 3 across the time-series data. This layer extracts local spatial structures, filtering out noise and detecting sudden, sharp micro-bursts to create a condensed feature map. This map is subsequently fed into the Bi-LSTM layer. Unlike a standard Long Short-Term Memory (LSTM) that processes data in one direction, our Bi-LSTM evaluates the sequence in both forward and backward directions simultaneously.
This dual-processing maximizes the contextual understanding of the data flow, enabling the model to learn long-term temporal dependencies and accurately predict abrupt load transitions.

To execute the predictions, we employ a multi-step time series forecasting approach. For simulating continuous real-world network operations, we apply a ``Stitched Continuous Forecasting'' method, iteratively advancing 10 slots at a time to generate an unbroken forecast trajectory.

The Bi-LSTM framework does not perform the bandwidth allocation itself. It concludes the forecasting stage by providing the predictive intelligence required for dynamic scheduling. This predictive foundation overcomes the limitations of purely reactive schedulers that suffer from increased latency when data queues unexpectedly accumulate. By anticipating incoming traffic bursts in advance, the predictions are fed directly into the RL framework, enabling the agent to proactively optimize the SBFD frequency split ratio.

\subsection{RL Approach}

The proposed work models dynamic SBFD scheduling as an MDP \cite{bellman1957markovian}, employing a DDQN agent for slot-by-slot decision-making. While standard Deep Q-Network (DQN) effectively handles complex traffic states, it relies on a single network for both action selection and evaluation, which often causes overestimation bias.
DDQN resolves this by separating the processes: an Online Network analyzes the 22-dimensional state to select the optimal action, while an independent Target Network evaluates that action, ensuring a stable allocation policy.

After validation of the Bi-LSTM module, the dynamic resource allocation problem is solved using the DDQN agent. The environment for this DDQN-based scheduler is characterized by three components: the state representation, the action space, and the reward formulation.

\textbf{State Representation:} The state vector has 22 dimensions. The first 20 values are the Bi-LSTM output, i.e., predictions for 10 future UL and 10 future DL traffic slots. The remaining 2 values are the current UL and DL queue sizes, normalized by dividing by 10 times the link capacity. This state design gives the agent both future and current traffic status and allows proactive decisions.

\textbf{Action Space:}  The action space is defined as a discrete set of UL/DL allocation ratios, allowing the agent to adapt the SBFD transmission pattern in response to predicted traffic and instantaneous congestion. Instead of selecting a specific frame pattern, the agent selects a precise resource split ratio between UL and DL, with the strict rule that the DL should never degrade below a minimum value.

\textbf{Reward and Training:} After the agent chooses a split, the system calculates a reward based on the actual traffic demand. It assigns higher rewards for efficient data delivery and applies penalties when queues build up. Through training, the agent learns to divide the available bandwidth between UL and DL based on the current traffic conditions. The scheduler then serves the data according to this split. Any unserved data remains in the queue for the next time slot, which mimics how real base stations handle buffered packets rather than simply discarding them.

By combining these two models, the system can both understand the traffic behavior and act in real time. The Bi-LSTM provides the intelligence to read the traffic, and the DDQN provides the ability to make smart allocations based on that reading.

\section{Methodology and Evaluation}

\subsection{Methodology}
We use a dataset that contains 1,000,000 network traffic records with two features: UL and DL traffic, both measured in bits per slot. Before feeding it to the model, both features are normalized to the 0-1 range using min-max scaling. The normalized data are then split into 3 categories: 80\% for training (800,000 samples), 10\% for validation (100,000 samples), and 10\% for testing (100,000 samples). The data is restructured using a sliding window technique.
In this setup, the algorithm takes a continuous history of 30 past time slots as its input to simultaneously forecast the traffic demands for the upcoming 10 time slots.

This structured sequence is processed by the hybrid forecasting architecture. The input first passes through a 1D-CNN layer equipped with 64 filters and a kernel size of 3. The output is then refined using a Batch Normalization layer and a MaxPooling layer (with a pool size of 2). A 20\% dropout layer is applied immediately after for regularization. Next, a 128-unit Bi-LSTM layer extracts temporal dependencies, followed by an additional 30\% dropout layer. Finally, a fully connected Dense layer outputs the raw forecasted values, which are reorganized by a Reshape layer into a $10 \times 2$ matrix.

The network is trained using the Adam optimization algorithm \cite{kingma2015adam} 
alongside the Huber loss function, making the system highly resilient to the sudden 
traffic outliers and extreme bursts typical in 6G wireless communication.
 The training process feeds the data to the model in batches of 32 samples and evaluates performance using the Mean Absolute Error (MAE) metric. To maximize efficiency, the training is scheduled for 5 epochs and utilizes two automated monitoring systems. A Model Checkpoint automatically saves the network weights whenever the validation loss reaches a new low. An Early Stopping mechanism with a patience of 4 actively monitors the validation loss; if the model's accuracy fails to improve for four consecutive steps, the training halts early and automatically restores the weights with the best performance. This ensures the final forecasting model is both highly accurate and well generalised for unseen traffic data.

Following the traffic prediction, the DDQN agent observes the 22-dimensional state space to dynamically manage the SBFD network resources. Based on this state, the agent selects an action from five discrete bandwidth split ratios. These choices dictate the exact proportion of network capacity allocated to UL versus DL, ranging sequentially from a 40:60 split down to a strictly DL-focused 0:100 split.

Once a bandwidth split is chosen, the environment evaluates the decision by calculating a reward based on queue management efficiency. The core reward is derived from the geometric mean of the UL and DL satisfaction rates, which measure the fraction of successfully served data against the total data waiting in the queues. Specific numerical penalties are subtracted from this base reward. The algorithm penalises the agent for wasting spectrum on empty queues, failing to allocate bandwidth to bursting queues that exceed capacity thresholds, and unnecessarily switching bandwidth ratios between consecutive steps. This targeted penalty system trains the agent to handle sudden traffic bursts without causing disruptive network jitter.

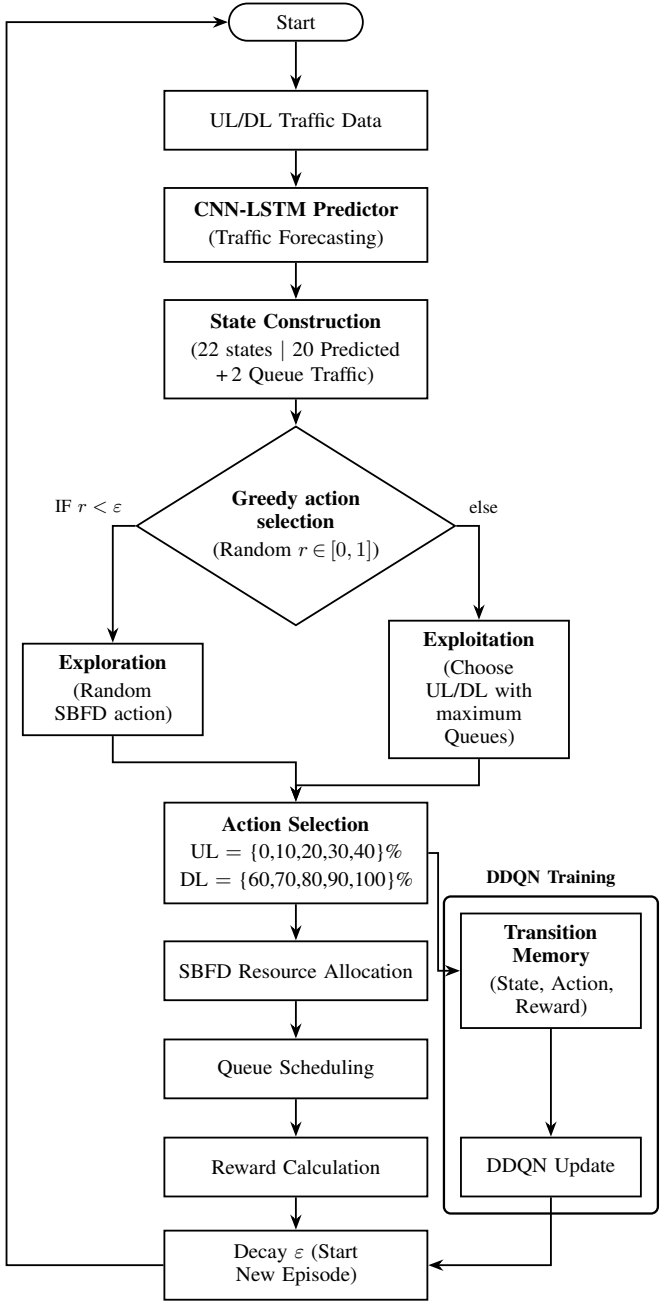
\begin{figure}[htbp]
\centering
\resizebox{\columnwidth}{!}{
\begin{tikzpicture}[
  >={Stealth[length=5pt,width=4pt]},
  line width=0.8pt,
  proc/.style = {rectangle, draw, text width=3.6cm, align=center,
                 inner sep=6pt, font=\small, minimum height=0.9cm},
  side/.style = {rectangle, draw, text width=2.4cm, align=center,
                 inner sep=5pt, font=\small, minimum height=0.9cm},
  dec/.style  = {diamond, draw, text width=2.8cm, aspect=1.6,
                 align=center, inner sep=2pt, font=\small},
  arr/.style  = {->, >=Stealth, line width=0.8pt}
]
 
\node[proc] (traf)   at (0, -1.5) {UL/DL Traffic Data};
\node[proc] (cnn)    at (0, -3.1) {\textbf{CNN-LSTM Predictor}\\[2pt]
                                    (Traffic Forecasting)};
\node[proc] (state)  at (0, -5.0) {\textbf{State Construction}\\[2pt]
                                    (22 states $|$ 20 Predicted\\[1pt] +\,2 Queue Traffic)};
\node[dec]  (greedy) at (0, -7.7) {\textbf{Greedy action selection}\\[2pt]
                                    (Random $r\!\in\![0,1]$)};
 
\node[side] (expl) at (-2.8, -10.2) {\textbf{Exploration}\\[2pt]
                                     (Random SBFD action)};
\node[side] (expt) at ( 2.8, -10.2) {\textbf{Exploitation}\\[2pt]
                                     (Choose UL/DL with\\ maximum Queues)};
 
\node[proc] (action) at (0,-12.7) {\textbf{Action Selection}\\[2pt]
                                    UL $=$ \{0,10,20,30,40\}\%\\[2pt]
                                    DL $=$ \{60,70,80,90,100\}\%};
\node[proc] (sbfd)   at (0,-14.5) {SBFD Resource Allocation};
\node[proc] (qsched) at (0,-16.0) {Queue Scheduling};
\node[proc] (rwrd)   at (0,-17.5) {Reward Calculation};
\node[proc] (decay)  at (0,-19.0) {Decay $\varepsilon$ (Start New Episode)};
 
\node[side] (tmem) at (3.9,-14.5) {\textbf{Transition Memory}\\[2pt]
                                    (State, Action, Reward)};
\node[side] (dupd)   at (3.9,-17.5) {DDQN Update};
 
\node[rounded rectangle, draw, font=\small,
      minimum width=2.4cm, inner sep=5pt]
     (start) at (0, 0) {Start};
 
\begin{scope}[on background layer]
  \node[rectangle, draw, rounded corners=3pt, line width=1.0pt,
        fit=(tmem)(dupd), inner sep=7pt,
        label={[font=\footnotesize\bfseries]above:DDQN Training}] {};
\end{scope}
 
\draw[arr] (start.south)  -- (traf.north);
\draw[arr] (traf.south)   -- (cnn.north);
\draw[arr] (cnn.south)    -- (state.north);
\draw[arr] (state.south)  -- (greedy.north);
 
\draw[arr] (greedy.west) -| node[pos=0, above left=2pt, font=\footnotesize]{IF $r<\varepsilon$} (expl.north);
 
\draw[arr] (greedy.east) -| node[pos=0, above right=2pt, font=\footnotesize]{else} (expt.north);
 
\draw[arr] (expl.south) -- ++(0,-0.4) -| (action.north);
\draw[arr] (expt.south) -- ++(0,-0.4) -| (action.north);
 
\draw[arr] (action.south) -- (sbfd.north);
\draw[arr] (sbfd.south)   -- (qsched.north);
\draw[arr] (qsched.south) -- (rwrd.north);
\draw[arr] (rwrd.south)   -- (decay.north);
 
\draw[arr] (action.east) -- ++(0.15,0) |- (tmem.west);
 
\draw[arr] (tmem.south) -- (dupd.north);
 
\draw[arr] (dupd.south) |- (decay.east);
 
\draw[arr] (decay.west) -- ++(-2.4,0) |- (start.west);
 
\end{tikzpicture}
}
\caption{Proposed Framework for traffic prediction and dynamic Sub-band allocation.}
\label{fig:allocation_flowchart}
\end{figure}

The internal DDQN architecture consists of a main network and a separate target network, both built using two hidden dense layers of 64 neurons each. The agent navigates its environment using an epsilon-greedy exploration strategy, starting with completely random actions and gradually decaying this randomness to rely on its learned optimal policies. During operation, every interaction (the initial state, chosen action, resulting reward, and the subsequent state) is saved into an experience replay memory buffer capable of storing up to 100,000 past events. To continuously improve its decision-making, the agent randomly samples mini-batches from this memory. The decoupled selection-evaluation mechanism described in Section III-B is optimized using the Mean Squared Error loss function, ensuring a reliable resource allocation strategy. The flow chart in Fig.~\ref{fig:allocation_flowchart} describes the overall algorithm.

\subsection{Evaluation Results}
\begin{figure}[!t]
    \centering
    \includegraphics[width=\columnwidth]{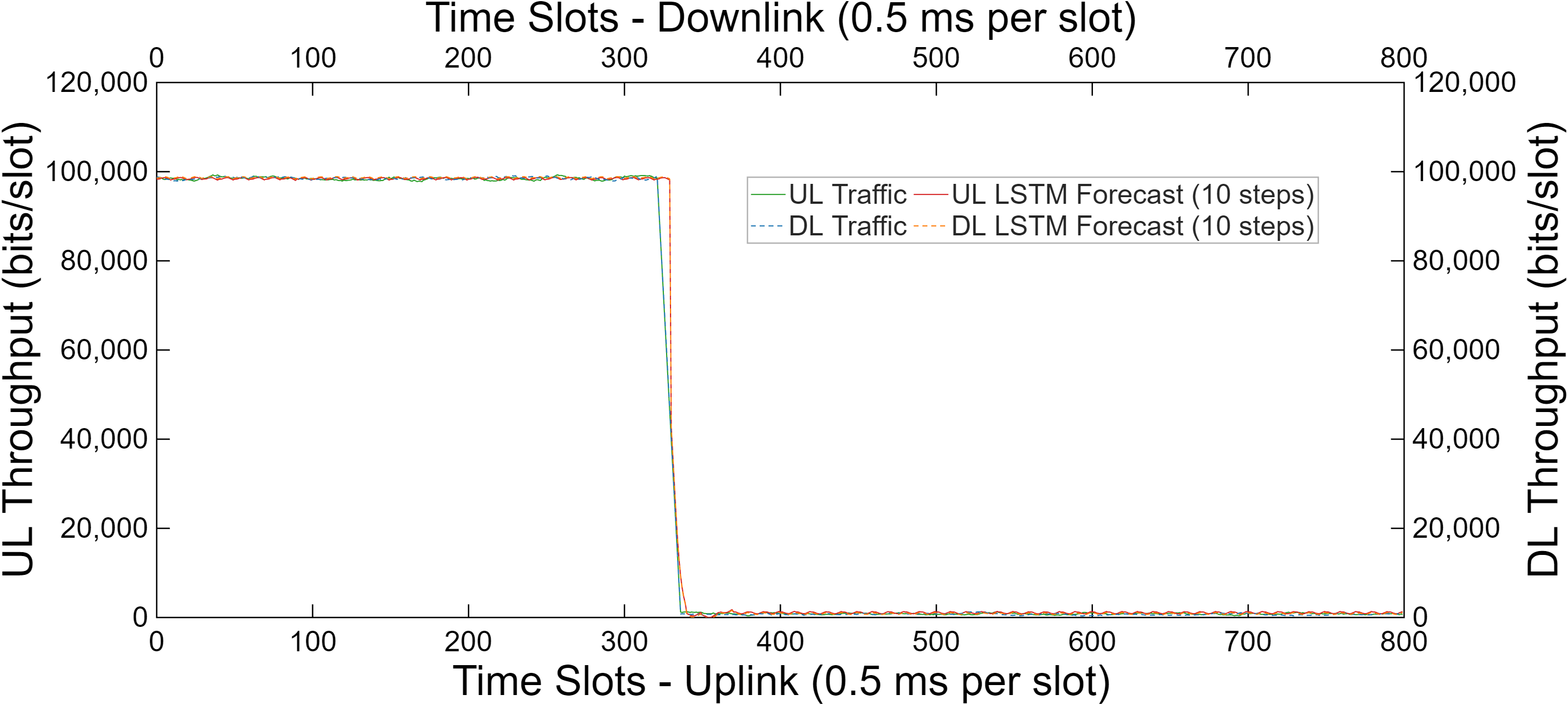}
    \caption{Bi-LSTM traffic prediction performance for UL and DL over a 10-slot horizon.}
    \label{fig:lstm_prediction}
\end{figure}

To validate the proposed models, we used the same MMPP dataset \cite{el2024markov} discussed earlier. After training, the Bi-LSTM iteratively predicted future slots of UL and DL traffic for a duration of up to 800 slots. The predictions and real traffic almost converged. Fig.~\ref{fig:lstm_prediction} visualizes the UL and DL traffic prediction stitched together across time. At time $t$, the model predicts slots $t$ to $t{+}9$, then shifts to $t{+}10$ and predicts slots $t{+}10$ to $t{+}19$, continuing iteratively to produce a single unbroken forecast line.

The UL MAE is 969 bits/slot (0.97\% of the 100,000 bits/slot capacity), and DL MAE is 849 bits/slot (0.85\% of capacity). The stitched predictions confirm that the Bi-LSTM maintains long-term phase alignment without drifting and recovers from transient errors, keeping the average prediction error below 10\% over extended periods.

\begin{figure}[!t]
    \centering
    \includegraphics[trim={0cm 0cm 0cm 0cm}, clip, width=\columnwidth]{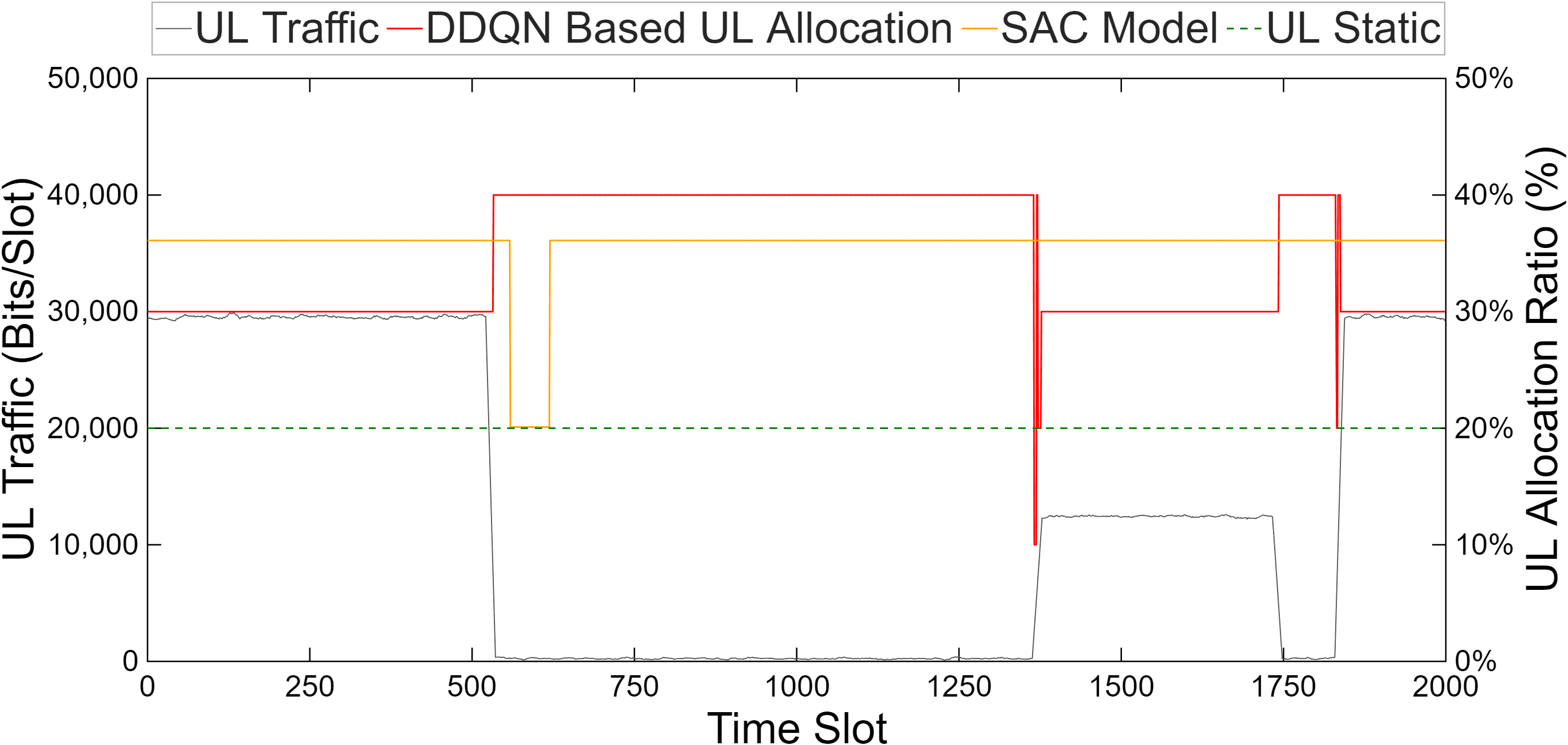}
    \caption{Dynamic Allocation based on UL demand in DDQN Model vs SAC-Discrete model in SBFD slots}
    \label{fig:ul_allocation_sac}
\end{figure}

To evaluate the RL agent, UL and DL demands were plotted alongside the agent's allocation ratios. Fig.~\ref{fig:ul_allocation_sac} shows the UL demand and the corresponding UL allocation by the DDQN agent. During the first 500 slots, UL demand peaks (~30,000 bits/slot) and the agent maintains an allocation of 30\%. Between slots 500 and 1380, UL demand drops to near zero; the agent increases UL allocation to 40\% to maintain a balanced split. From slot 1380 to 1750, a moderate UL surge (~12,000 bits/slot) appears, and the agent reduces UL allocation back to 30\%.

\begin{figure}[!t]
    \centering
    \includegraphics[trim={0cm 0cm 0cm 0cm}, clip, width=\columnwidth]{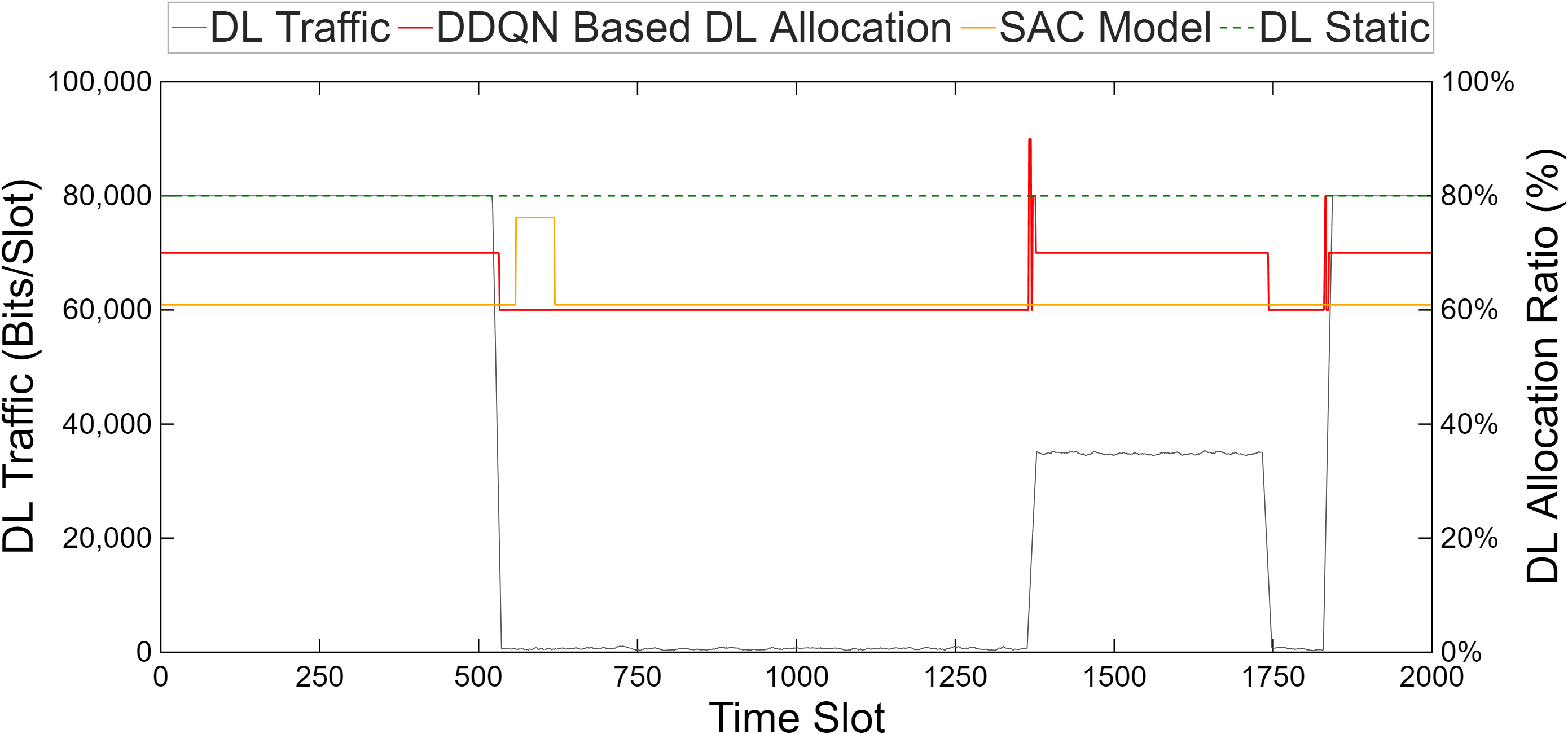}
    \caption{Dynamic Allocation on DL demand in DDQN model vs SAC-Discrete model in SBFD slots}
    \label{fig:dl_allocation_sac}
\end{figure}

Fig.~\ref{fig:dl_allocation_sac} shows DL allocation corresponding to DL demand. During slots 0 to 500, DL demand equals 80,000 bits/slot and the agent holds DL allocation at 70\%, with the remaining 30\% reserved for UL. Near slot 1380, UL demand jumps to ~12,000 bits/slot and DL to ~35,000 bits/slot. During the preceding idle period, DL was held at 60\%; when demand returns, the agent immediately raises DL allocation back to 70\%. Between slots 1400 to 1750, DL demand settles around 35,000 bits/slot and the agent maintains a consistent 70\% allocation.

Around slot 500, both demands drop to near zero, yet the agent decreases DL allocation to 60\%. This is not a failure; when no traffic arrives in either direction, any action switch incurs a switching penalty without throughput gain. Since the baseline traffic is roughly a 1:1 ratio, the agent defaults to a balanced split (60/40), minimizing penalties and decreasing wastage of the DL spectrum. This behavior persists through all subsequent dead periods (slots 600 to 1350 and the later idle window). When both demands return near slot 1800, the agent promptly re-adjusts DL to 70\%, confirming instant recovery from idle-state behavior.

The static baseline, shown in both figures, cannot adapt to demand changes and wastes resources. The DDQN agent avoids rapid switching when both demands are high and holds a stable configuration during idle periods, demonstrating demand-aware behavior suited to bursty, asymmetric SBFD environments.

We compared our Bi-LSTM+DDQN scheduler against the SAC-Discrete scheduler proposed in \cite{11174344}. In \cite{11174344}, the authors use a geometric mean reward of cumulative UL and DL throughput for proportional fairness, selecting between TDD and SBFD frames. We implemented their SAC-Discrete algorithm and trained it on our MMPP dataset. For fair comparison, we restricted the SAC-Discrete agent to the 3 SBFD frames from their work: XXXXX, XXXXU, and DXXXU, and computed the UL/DL capacity of each frame from the resource block specifications in \cite{11174344}.

A key difference is that the SAC-Discrete environment has no queues, and unserved data at each slot is permanently lost. In \cite{11174344}, this limitation does not apply because their evaluation uses a 3GPP-compliant system-level simulator with packet queues, HARQ retransmissions, and buffer management.

Over 100,000 slots, the SAC-Discrete agent selected Action~1 (XXXXU: 36.1\% UL / 60.9\% DL) in 98,424 slots (98.4\%), Action~0 (XXXXX: 20.1\% UL / 76.2\% DL) in 1,576 slots (1.6\%), and Action~2 (DXXXU: 32.1\% UL / 65.7\% DL) zero times. This extreme convergence indicates a fixed policy that fails to dynamically adjust the subband split.

In Fig.~\ref{fig:dl_allocation_sac}, the SAC-Discrete agent holds constant DL allocation at 60.9\%, while our DDQN agent varies between 60\% and 70\%. During peak DL demand (80,000 bits/slot, slots 0 to 500), the SAC-Discrete's 60.9\% falls short of both the static baseline's 80\% and the DDQN's 70\%. During dead periods, the DDQN adjusts DL to 60\%, while the SAC-Discrete remains at 60.9\%.

The reason the SAC-Discrete agent locks onto a single frame is its geometric mean reward, which is sensitive to the smaller component. Since UL demand averages roughly three times less than DL, the agent identifies UL starvation as the greatest risk. XXXXU offers the highest UL capacity (36.1\%), making it the safest choice. Without queues, unserved data is lost each slot, so the agent is never penalized for over-allocating to an idle direction and has no incentive to switch frames during idle periods. Action~2 (DXXXU: 32.1\% UL) offers slightly less UL than XXXXU with only marginal DL gain, making it suboptimal under the geometric mean objective.

\begin{table}[!t]
\centering
\caption{Comparison of Proposed Bi-LSTM + DDQN and SAC-Discrete Approaches}
\label{tab: comparison}
\renewcommand{\arraystretch}{1.3}
\begin{tabular}{|p{2.0cm}|p{3.0cm}|p{2.5cm}|}
\hline
\textbf{Feature} & \textbf{Bi-LSTM + DDQN (Proposed)} & \textbf{SAC-Discrete \cite{11174344}} \\
\hline
\textbf{RL Algorithm} & Double Deep Q-Network (DDQN) & Soft Actor-Critic Discrete (SAC-D) \\
\hline
\textbf{Traffic Predictor} & Bi-LSTM (predicts next 10 slots) & None (uses instantaneous demand) \\
\hline
\textbf{Dataset} & MMPP  & MMPP  \\
\hline

\textbf{Frame Type} & SBFD only & SBFD only (\cite{11174344} uses TDD + SBFD) \\
\hline
\textbf{Action Space} & 5 (UL: 40\%, 30\%, 20\%, 10\%, 0\%) & 3 (XXXXX: 20.2\% UL, XXXXU: 36.1\% UL, DXXXU: 32.1\% UL) \\
\hline
\textbf{States} & Bi-LSTM predictions (20) + queue sizes (2) = 22 dimensions & Instantaneous UL/DL demand + cumulative throughput \\
\hline
\textbf{Reward Function} & Geometric mean of UL/DL satisfaction $-$ conditional queue \& switching penalties & Incremental geometric mean of cumulative UL \& DL \\

\hline
\textbf{Queue} & Yes (unserved traffic is buffered) & No (unserved traffic is lost) \\
\hline

\textbf{Allocation Behavior} & Dynamic: UL varies 10\% to 40\%, DL varies 60\% to 90\% based on real-time traffic demand & Fixed: converges to a single frame (XXXXU, 36.1\% UL/ 60.9\% DL) in 98.4\% of slots \\
\hline

\end{tabular}
\end{table}

Table \ref{tab: comparison} summarizes the architectural and behavioral differences between our proposed model and the SAC-Discrete baseline \cite{11174344}. The inability of the SAC-Discrete agent to perform dynamic allocation stems from its lack of traffic forecasting and the absence of stateful queues in its environment. Without queue accumulation to penalize poor resource allocation during bursts or idle periods, the SAC-Discrete agent safely defaults to a single, statically optimal frame configuration. In contrast, the queue-based feedback and proactive 10-slot lookahead force our DDQN agent to develop a genuinely dynamic, demand-aware policy.

The performance gains of the DDQN agent over both baselines are as follows: On the UL side, the static baseline fixes UL at 20\% (20,000~bits/slot). During peak UL demand of 30,000~bits/slot, this leaves a shortfall of 10,000~bits/slot per slot and queues grow continuously. The DDQN agent, using the Bi-LSTM forecast, raises UL allocation to 30\% (30,000~bits/slot), which exactly matches the demand and keeps the queue empty. This is a 100\% reduction in UL queue buildup at peak load, and more than 90\% overall across all active traffic periods compared to the static baseline.

On the DL side, during peak demand of 80,000~bits/slot (slots 0--500), the DDQN allocates 70\% of the 100,000~bits/slot link capacity, giving 70,000~bits/slot for DL. The SAC-Discrete stays fixed at 60.9\%, giving only 60,900~bits/slot. This is a difference of 9,100~bits/slot, which is a DL throughput gain of about 14.9\% for the DDQN over the SAC-Discrete. The static baseline gives 80\% to DL but only 20\% to UL. Since peak UL demand is 30,000~bits/slot and the static provides only 20,000~bits/slot, peak UL utilization under the static is just 66.7\%, and queue buildup persists. The DDQN avoids this by matching its UL allocation to the actual demand, achieving 100\% peak UL utilization while also providing 9.1 percentage points more DL capacity than the SAC-Discrete in every active period.

\section{Conclusion}

In this paper, we introduced a dynamic sub-band allocation method for SBFD systems using a combined Bi-LSTM and DDQN approach. The Bi-LSTM acts as a predictor that forecasts upcoming UL and DL traffic, providing the DDQN agent with the exact information it needs to adjust resource split ratios dynamically. Legacy static allocation methods and models from existing literature like the SAC-Discrete scheduler \cite{11174344} waste the available spectrum because they keep the resource split almost fixed, ignoring the actual traffic demand. Instead of depending on these static or semi-static policies, our model constantly adapts based on real-time network conditions. The simulation results confirm that our approach tracks rapid traffic changes much better than the static allocation and the SAC-Discrete model, which struggles to adjust to dynamic loads. During peak traffic bursts, the proposed method reduces the buildup of the UL queue by more than 90\% compared to the static baseline. The SAC-Discrete scheduler sacrifices approximately 9.1 percentage points of DL capacity by locking its allocation at 60.9\% regardless of demand, while our DDQN dynamically raises DL to 70\%, delivering 9,100 more bits/slot to DL in every active period. This adaptation yields 100\% peak UL spectrum utilization and improves DL throughput by approximately 15\% (14.9\%) relative to the SAC-Discrete baseline. By maintaining stable split ratios during idle periods, the model also minimizes switching overhead. The system maximizes spectrum efficiency by efficiently allocating total bandwidth based on traffic demand. Future work will focus on testing the model using actual network traffic, with multiple cells and base stations, and deploying reinforcement learning with multiple agents.

\bibliographystyle{IEEEtran}
\bibliography{reference}

\end{document}